\documentclass[aps,twocolumn,prb,preprintnumbers,amsmath,amssymb,superscriptaddress,floats]{revtex4}

\usepackage{graphicx}
\usepackage{bm}
\usepackage{hyperref}
\usepackage{natbib}
\usepackage{braket}
\usepackage{float}
\usepackage{color}

\begin{document}

\title{Anomalous 3D bulk AC conduction within the Kondo gap of SmB$_6$ single crystals}

\author{N. J. Laurita}
\affiliation{The Institute for Quantum Matter, Department of Physics and Astronomy, The Johns Hopkins University, Baltimore, MD 21218, USA}

\author{C. M. Morris}
\affiliation{The Institute for Quantum Matter, Department of Physics and Astronomy, The Johns Hopkins University, Baltimore, MD 21218, USA}

\author{S. M. Koohpayeh}
\affiliation{The Institute for Quantum Matter, Department of Physics and Astronomy, The Johns Hopkins University, Baltimore, MD 21218, USA}

\author{P. F. S. Rosa}
\affiliation{Los Alamos National Laboratory, Los Alamos, NM 87545, USA} 
\affiliation{Department of Physics and Astronomy, University of California, Irvine, CA 92697, USA}

\author{W. A. Phelan}
\affiliation{The Institute for Quantum Matter, Department of Physics and Astronomy, The Johns Hopkins University, Baltimore, MD 21218, USA}
\affiliation{Department of Chemistry, Johns Hopkins University, Baltimore, Maryland, 21218, USA}

\author{Z. Fisk}
\affiliation{Department of Physics and Astronomy, University of California, Irvine, CA 92697, USA}

\author{T. M. McQueen}
\affiliation{The Institute for Quantum Matter, Department of Physics and Astronomy, The Johns Hopkins University, Baltimore, MD 21218, USA} 
\affiliation{Department of Chemistry, Johns Hopkins University, Baltimore, Maryland, 21218, USA}

\author{N. P. Armitage}
\affiliation{The Institute for Quantum Matter, Department of Physics and Astronomy, The Johns Hopkins University, Baltimore, MD 21218, USA}

\date{\today}

\begin{abstract}
\noindent The Kondo insulator SmB$_6$ has long been known to display anomalous transport behavior at low temperatures, T $<5$ K.  In this temperatures range, a plateau is observed in the dc resistivity, contrary to the exponential divergence expected for a gapped system. Recent theoretical calculations suggest that SmB$_6$ may be the first topological Kondo insulator (TKI) and propose that the residual conductivity is due to topological surface states which reside within the Kondo gap.  Since the TKI prediction many experiments have claimed to observe high mobility surface states within a perfectly insulating hybridization gap.  Here, we investigate the low energy optical conductivity within the hybridization gap of single crystals of SmB$_6$ via time domain terahertz spectroscopy.  Samples grown by both optical floating zone and aluminum flux methods are investigated to probe for differences originating from sample growth techniques.  We find that both samples display significant 3D bulk conduction originating within the Kondo gap.  Although SmB$_6$ may be a bulk dc insulator, it shows significant bulk ac conduction that is many orders of magnitude larger than any known impurity band conduction.   The nature of these in-gap states and their coupling with the low energy spin excitons of SmB$_6$ is discussed.  Additionally, the well defined conduction path geometry of our optical experiments allows us to show that any surface states, which lie below our detection threshold if present, must have a sheet resistance of R$/\square \ge$ 1000 $\Omega$.
\end{abstract}

\maketitle

\section{Introduction}
Topological states of matter have dominated the condensed matter research landscape in recent years and none more so than topological insulators.  Topological insulators possess bulk band inversion due to strong spin-orbit coupling resulting in chiral spin-momentum locked surface states, which are protected by time reversal or crystal symmetry \cite{Fu2007, Hasan2010, Moore2007, Qi2011, Roy2009, Fu2011}.  Since their prediction, a plethora of experimental evidence has corroborated their existence and investigated the corresponding physics \cite{Hsieh2008, Zhang2009, TongZhang2009, Okada2011, Aguilar2012}. However, this class of topological insulators are in their essence non-interacting systems. Additionally, clean samples with the Fermi energy deep within the bulk insulating gap have proven challenging to synthesize, limiting their potential applications.  Merging strong electron correlations with non-trivial topology is an exciting avenue to pursue exotic many-body quantum ground states with a truly insulating bulk.

The Kondo insulator SmB$_6$, sometimes referred to as a mixed-valent semiconductor \cite{Varma1976}, has recently been proposed as such a correlated, topologically non-trivial state \cite{Dzero2010, Takimoto2011, Dzero2012, Fu2013, Alexandrov2013}.  SmB$_6$ undergoes a crossover from metal to insulator behavior with reducing temperature which can be attributed to the opening of a bulk band gap of $\Delta _\text{K}$ $\approx$ 15-20 meV.  The gap is believed to originate from hybridization between localized 4$f$ electrons near the Fermi level and itinerant 5$d$ electrons \cite{Coqblin1968, Mott1974, Varma1976}.  Correspondingly, the dc resistivity shows an exponential divergence with reducing temperature, as expected for a gapped system, but then surprisingly plateaus at temperatures T $<$ 5K, suggesting an additional conduction mechanism \cite{Menth1969, Nickerson1971}.  Although first interpreted as stemming from impurity states \cite{Kebede1996, Gabani2001}, the low temperature resistivity plateau has recently been proposed to arise from topological surface states residing within the Kondo gap, suggesting SmB$_6$ to be the first example of a topological Kondo insulator (TKI) \cite{Dzero2010, Takimoto2011, Dzero2012, Fu2013, Alexandrov2013}.  Non-trivial topology is supported by recent calculations which propose SmB$_6$ possesses three Dirac cones located at high symmetry points of the Brillouin zone \cite{Fu2013, Takimoto2011, Alexandrov2013}.

Since the TKI prediction of SmB$_6$, experimental evidence of surface conduction at low temperatures has been reported via transport \cite{Kim2013, Kim2014, Wolgast2013, Chen2015} and tunneling spectroscopy \cite{XZhang2013, Yee2013}.  Meanwhile other techniques such as torque magnetometry \cite{Li2014}, photoemission \cite{Nuepane2013, Miyazaki2012, Xu2014, Jiang2013, Frant2013}, and neutron scattering \cite{Fuhrman2015} also report findings consistent with the TKI prediction.  This has led many to hail SmB$_6$ as the quintessential TKI, with high mobility surface states wrapping a perfectly insulating bulk.  

These claims, however, should be considered in conjunction with previous low energy ac optical conductivity experiments of SmB$_6$ single crystals which have claimed evidence for \textit{localized states} within the Kondo gap at the lowest temperatures and ac conductivities orders of magnitude higher than the dc value \cite{Travaglini1984, Jackson1984, Kimura1994, Nanba1993, Gorshunov1999}. These observed localized states are in stark contrast to the expected Drude response, indicative of free charge carries, observed from the surface states of Bi$_2$Se$_3$ \cite{Aguilar2012, Wu2013}.  However, these optical experiments on SmB$_6$ single crystals pre-date the TKI prediction and may require reinterpretation.  Additionally, results from a number of heat capacity experiments reveal a very large low temperature fermionic heat capacity with a $\gamma$ coefficient that is 2-25 mJ$/$mol $\cdot$K$^2$ (the same as some correlated metals) which has been shown to be of bulk origin \cite{Wakeham2016} and therefore seemingly at odds with a bulk gapped state \cite{Flachbart2006, Gabani2001, Phelan2014}.

While the origin of these in-gap states remains an open question, recent experiments suggest that impurities and disorder do play an important role in the low temperature physical properties of SmB$_6$, perhaps even in the topological aspects.  Phelan \textit{et al.} \cite{Phelan2014} demonstrated that the low temperature resistivity plateau can be tuned as a function of carbon or aluminum doping, typical impurities found in SmB$_6$ depending on the synthesis method and quality of seed materials.  The effects of disorder in the form of Sm$^{+2,3}$ vacancies have also been examined and shown to produce significant changes in the low temperature physical properties of SmB$_6$ \cite{Phelan2016, Valentine2016}.  Recent Raman spectroscopy measurements show that Sm$^{+2,3}$ vacancies on the order of only 1\% can effectively close the bulk gap  \cite{Valentine2016}. In this regard, recent theoretical calculations predict the topological properties of SmB$_6$ to be strongly dependent on Sm$^{+2,3}$ valence \cite{Alexandrov2013}, which will correlate with sample imperfections.  These results suggest that synthesis method, impurity concentration, and disorder are important considerations and warrant further investigation in SmB$_ 6$.


Low energy optical experiments are well suited for investigating the in-gap conduction in SmB$_6$.  Additionally, transmission experiments performed as a function of sample thickness can separate surface and bulk conduction and have therefore been successful in the field of topological insulators \cite{Aguilar2012, Wu2013}. However, optical transmission experiments on the rare-earth hexaborides can be exceptionally challenging due to their unusually large index of refraction.  Moreover, as we discuss below, SmB$_6$ itself has substantial ac conduction that precludes simple transmission experiments.  Therefore, the optical properties of the hexaborides have been traditionally studied via reflection techniques, \cite{Travaglini1984, Kimura1994, Nanba1993} which rely on a Kramers-Kronig transform to obtain the real and imaginary parts of the response and possess substantially less signal to noise than what can be achieved in modern phase sensitive transmission experiments.  Transmission experiments of SmB$_6$ in the far infra-red have been performed with success \cite{Gorshunov1999, Ohta1991} but a detailed temperature and thicknesses dependence of the optical conductivity has not been provided.  Moreover, the continuous wave nature of previously used techniques can give artifacts due to standing wave resonances in the optical apparatus.

In this paper we present a comprehensive high resolution study of the optical properties of the potential topological Kondo insulator SmB$_6$ in the terahertz frequency range.  As the gap energy, $\Delta _\text{K}$ $\approx$ 15-20 meV, is larger than our experimental energy range, $\hbar \omega$ $\approx$ 1-8 meV, we directly probe states within the bulk insulating gap via the optical conductivity.  We compare samples grown via both optical floating zone and aluminum flux methods to examine differences originating from sample preparation, but only minor differences are found.  Transmission experiments performed as a function of sample thickness determine that the conduction of the in-gap states is predominantly 3D in nature.  Our results show that the ``perfectly insulating" bulk of SmB$_6$ in fact has significant 3D conduction at finite frequencies that is many orders of magnitude larger than any known impurity band conduction.  The potential origins of these states and their coupling to the low energy spin excitons of SmB$_6$ are discussed.  Additionally, the well defined conduction path geometry of our optical experiments allows us to place limits on the sheet resistance of potential surface states, which must lie below our detection threshold if present.

\section{Methods}
As stated in the introduction, the exceptionally high index of refraction of SmB$_6$, $n$ $\approx$ 25, in the THz regime presents experimental challenges for transmission measurements.  One can show from the Fresnel relations that the reflection coefficient of a sample with index of refraction $n$ = 25 at normal incidence is $r$ $\approx$ $[\frac{(n -1)}{(n+1)}]^2 \approx 85$\%.  Absorptions in the sample and reflection off the back surface drastically further reduce the transmission.  Therefore, novel methods for measuring SmB$_6$ single crystals are needed in order to achieve sufficient signal to noise.  

Correspondingly, we find SmB$_6$ samples are not sufficiently transmissive in the THz range until sample thicknesses of d $\leq$ 100 $\mu$m.  In order to achieve such thicknesses, SmB$_6$ samples were first double sided polished to a mirror finish to ensure plane parallel sides.  Samples were then mounted to a double side polished Al$_2$O$_3$ substrate of nominal thickness of 500 $\mu$m via mounting wax.  Once mounted, SmB$_6$ samples were not removed from the substrate for the remainder of the experiment.  Samples were then further polished to a thickness of 10's of $\mu$m as measured by a micrometer. Time domain terahertz (TDTS) transmission experiments were then performed.  The thickness dependent THz response of the samples was obtained by further polishing the samples in between TDTS measurements. 

TDTS transmission experiments were performed using a home built spectrometer within a temperature range of 1.6K to 300K \cite{Laurita2016}.  TDTS is a high resolution method for accurately measuring the electromagnetic response of a sample in the experimentally challenging THz range.  In a typical TDTS experiment, the electric field of a transmitted THz pulse through a sample mounted to a circular aperture is measured as a function of real time. Fourier transforming the measured electric field and referencing to an aperture of identical size allows access to the frequency dependent \textit{complex} transmission spectrum of the sample.  In this case the transmission is given by the expression

\begin{equation}
\widetilde{T} = \frac{4\widetilde{n}}{(\widetilde{n}+1)^2}e^{\frac{i \omega d}{c}(\widetilde{n}-1)}
\label{Transeq}
\end{equation}

\noindent where $d$ is the sample thickness, $\omega$ is the frequency, $c$ is the speed of light, $\widetilde{n}$ is the sample's complex index of refraction, and normal incidence has been assumed.  A Newton-Raphson based numerical inversion of the complex transmission is then used to obtain both the frequency dependent real and imaginary parts of the index of refraction.  In principle the index of refraction of the sample, $\widetilde{n} = n + ik = \sqrt{\epsilon \mu}$, contains both the electric and magnetic responses of the sample since THz fields can couple to both electric and magnetic dipole transitions.  However, we find the optical response of SmB$_6$ in the THz regime to have no magnetic field dependence in static fields H $\leq$ 7T and therefore assign all dissipation as stemming from electric effects.  This allows the complex optical conductivity to be determined from $n$ and $k$.

\begin{figure}
\includegraphics[width=1.0\columnwidth, height=2.75 in]{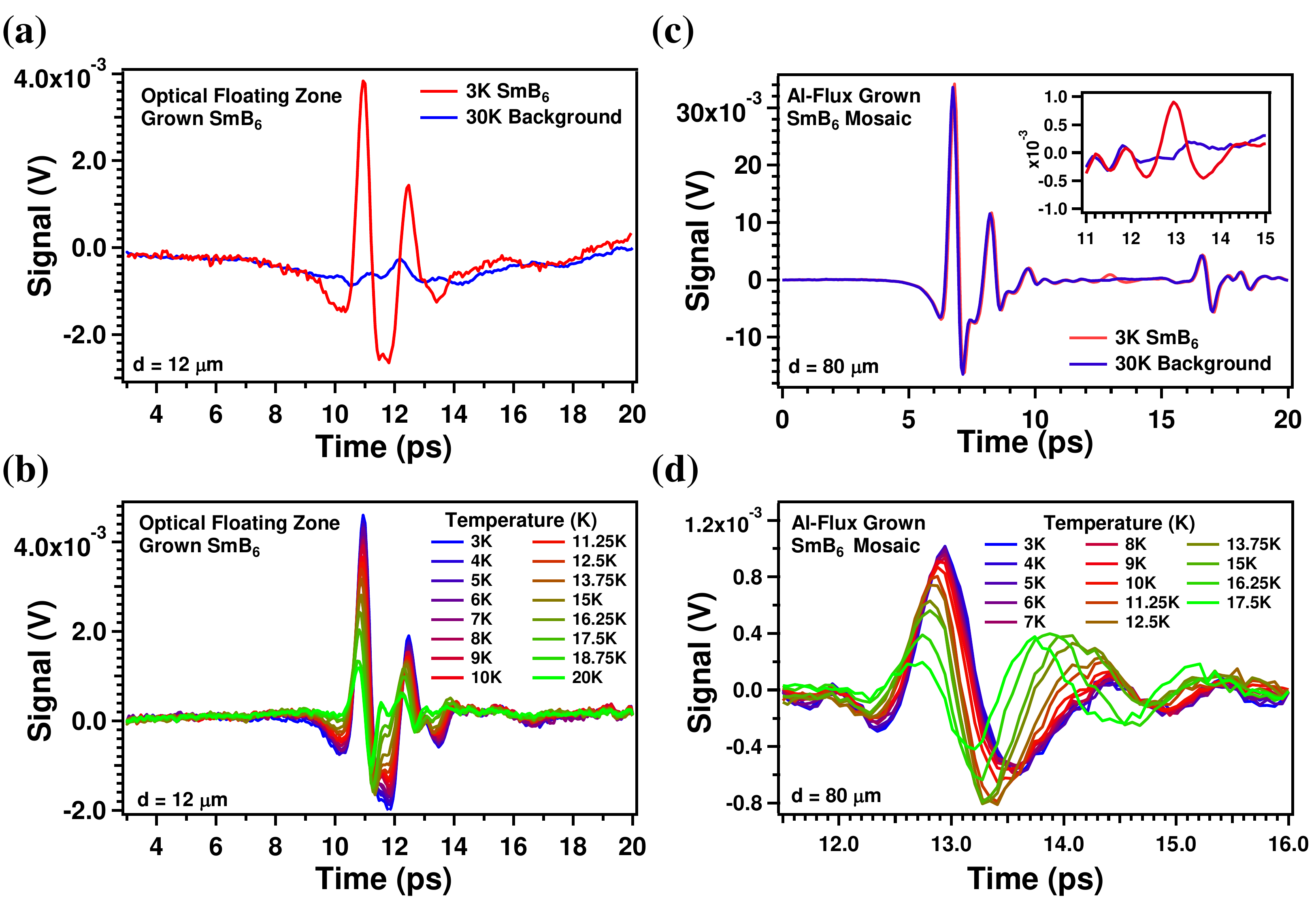}
\caption{(a) Transmitted electric field of an optical floating zone grown SmB$_6$ single crystal mounted to an Al$_2$O$_3$ substrate as a function of time at 3K and 30K.  The 30K signal is identified as background ``light leak" signal which is caused by diffraction of light around the sample. (b) Transmitted electric field as a function of temperature once the background signal shown in (a) is removed by subtraction.  (c) Transmitted electric field of the aluminum flux grown SmB$_6$ mosaic at 3K and 30K.  The light leak in this case is much larger due to diffraction between neighboring samples within the mosaic.  The two largest signals shown stem from light only transmitted through the Al$_2$O$_3$ substrate and must be subtracted from the reference substrate's transmitted electric field.  The inset shows an expanded view of the time window in which the signal from light transmitted through the SmB$_6$ mosaic is observed. (d) Transmitted electric field as a function of temperature once the background signal shown in (c) has been removed by subtraction. See text for more details.}
\label{Fig1}
\end{figure}

Mounting single crystals of SmB$_6$ to the Al$_2$O$_3$ substrates introduces a new interface which modifies the typical transmission expression presented in Eq. \ref{Transeq}.  In this case it is best to use an identical substrate as a reference as the transmission is then independent of the substrate's thickness.  For the case of a single crystal mounted to a substrate, Eq. \ref{Transeq} is modified as 

\begin{equation}
\widetilde{T} = \frac{2\widetilde{n}(\widetilde{n}_s + 1)}{(\widetilde{n}+1)(\widetilde{n}+\widetilde{n}_s)}e^{\frac{i \omega d}{c}(\widetilde{n}-1)}
\label{Transeq2}
\end{equation}

\noindent where $\widetilde{n}_s$ is the \textit{substrate's} complex index of refraction.  One can verify that in the case of no substrate, $\widetilde{n}_s$ = 1, Eq. \ref{Transeq} is recovered.  TDTS experiments were performed on the Al$_2$O$_3$ substrate with an aperture reference where it was found that $\widetilde{n}_s$ is well approximated by a real constant in the THz range, as expected for a good insulator, with a value of $n_s$ = 3.    

Long wavelength THz radiation restricts TDTS to samples with fairly large cross sectional areas.  Therefore, sample diameters greater than 3 mm are typically needed in order to achieve sufficient signal to noise.  Optical floating zone SmB$_6$ samples are therefore better suited for TDTS as single crystals can often be so large.  TDTS measurements on large floating zone crystals were performed on single crystal SmB$_6$ samples with the $\hat{c}$ [001] axis oriented out of the plane of the sample surface.  SmB$_6$ samples grown via the aluminum flux method are generally smaller than what is required for TDTS.  In order to achieve sufficient signal to noise on these samples, a ``mosaic" of 10 closely packed aluminum flux grown SmB$_6$ samples were mounted to an identical Al$_2$O$_3$ substrate.  The mosaic covered a rectangular spatial area of $\approx$ 3.5 mm $\times$ 6 mm in cross section.  All aluminum flux samples were oriented with the $\hat{c}$ [001] axis out of the plane of the sample surface. One should note that the cubic symmetry of the $Pm3m$ space group of SmB$_6$ ensures that the linear optical response is identical for incident THz $\vec{k}$ oriented along any of the principal axes of the crystal \cite{Armitage2014}.

Additional complications can arise in very low transmissivity samples as low absolute levels of incident radiation can - even for the single crystal -  be transmitted through cracks in the sample surface, through gaps in between single crystals mounted in the mosaic pattern, or around the cryostat itself.  We refer to this spurious signal as a ``light leak" and it must be removed from our data for accurate results.  Fig. \ref{Fig1} displays our methods for removing such light leak signal in the case of both single crystal optical floating zone samples and the aluminum flux grown mosaic, although both are qualitatively similar with the exception that the light leak is larger in the the case of the mosaic.  We find that even the thinnest SmB$_6$ samples become opaque to THz radiation at temperatures T $\approx$ 30K.  Presumably this stems from the bulk Kondo gap closing with increasing temperature.  Yet, a very small background light leak signal is still transmitted at and above 30K.  We identify this signal as the light leak as it is temperature independent from 30K to room temperature.  Additionally, we find our data are not systematic until this spurious signal is removed.  For the case of optical floating zone samples this signal is simply removed by subtracting the light leak signal as a function of time at T = 30K from the transmitted electric field of the sample at lower temperatures.  Fig. \ref{Fig1} (a) shows the transmitted THz electric field at 3K through an optical floating zone SmB$_6$ sample (d = 12 $\mu$m) as well as the 30K light leak signal.  Fig. \ref{Fig1} (b) shows the measured electric field of the same sample at temperatures below 30K once the light leak signal has been subtracted.

\begin{figure*}[tb]
\includegraphics[width=2.0\columnwidth, height=5in]{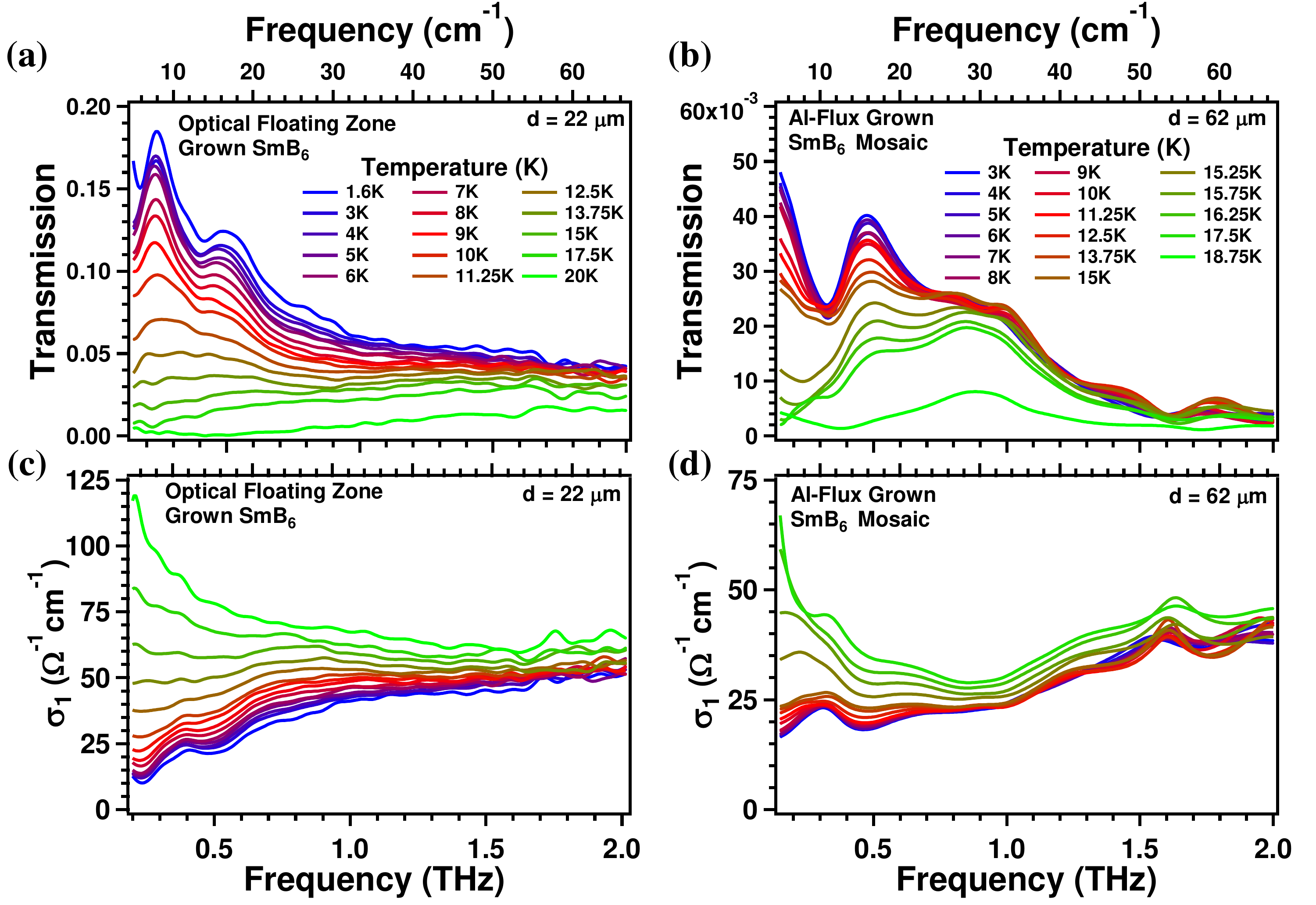}
\caption{(a,b) Magnitudes of the complex transmissions, as defined in Eq. \ref{Transeq2}, as a function of frequency and temperature for representative samples grown by both (a) optical floating zone and (b) aluminum flux methods. The two samples had thicknesses of 22 $\mu$m and 62 $\mu$m respectively. (c,d) Real part of the optical conductivity. $\sigma _1 (\omega, T)$, calculated from the transmissions shown in (a,b).}
\label{Fig2}
\end{figure*}

Removing the light leak signal from the aluminum flux grown mosaic contains an additional complication as the light leak in this case is much larger due to diffraction through spaces between neighboring samples of the mosaic.  It requires some additional considerations in analysis, that we believe are applied here for the first time.  In a similar manner as described above, the light leak signal as measured at 30K is subtracted from sample scans at lower temperatures.  However, the additional step of subtracting the light leak signal from the measured reference substrate's electric field is taken to ensure the transmission is accurate.  This step is not necessary for the optical floating zone samples as the light leak signal is substantially smaller than the transmitted substrate's electric field, $<$ 1\%.  However, the light leak is as large as 40\% for the SmB$_6$ mosaic. Fig \ref{Fig1} (c) shows the 3K and 30K measured electric field of the SmB$_6$ mosaic (d = 80 $\mu$m).  The first large pulse at $\approx$ 7 ps stems from light diffracting around and between neighboring samples of the mosaic and therefore only traveling through the Al$_2$O$_3$ substrate.  This signal is subtracted from the reference substrate's measured signal to correct the transmission.  The next largest signal at $\approx$ 17 ps is the first echo of light which has been reflected within the substrate twice.  The inset of the graph shows the signal in between these two substrate pulses where a small but finite signal of light transmitted through the SmB$_6$ mosaic can be seen at $\approx$ 13 ps.  Fig.  \ref{Fig1} (d) shows the extracted transmitted electric field of the SmB$_6$ mosaic as a function of temperature once the light leak has been subtracted.  

\section{Experimental Results}
\subsection{Low Energy Optical Response of SmB$_6$}

Figs. \ref{Fig2} (a,b) display the magnitude of the complex transmission, as defined in Eq. \ref{Transeq2}, as a function of temperature and frequency for two representative samples grown by optical floating zone (d = 22 $\mu$m) and aluminum flux methods (d = 62 $\mu$m) respectively.  Both samples show qualitatively similar behavior of the transmission.  At the lowest temperatures the largest transmission, $\approx$ 5 - 20 \% depending on sample thickness and synthesis method,  is observed at the lowest frequencies.  The transmission then quickly decreases with increasing frequency.  Although both samples show the same general features, we believe that the data for the floating zone crystal is more representative of the true spectral shape of  SmB$_6$ due to artifacts that can be introduced in the mosaic geometry.   For instance, we believe the dip in transmission of the aluminum flux grown mosaic sample at $\approx$ 0.3 THz is an artifact as it is not systematic between measurements and likely stems from imperfections in our method of removing the light leak signal as described above.  For both samples, the transmission gradually decreases with increasing temperature until becoming opaque in the THz range for T $\ge$ 30K for sample thicknesses d $>$ 10 $\mu$m.   As we will discuss below these features are generally consistent with residual conductivity within a gap which is closing or filling in with increasing temperature.

\begin{figure}
\includegraphics[width=1.0\columnwidth, height=5.0in]{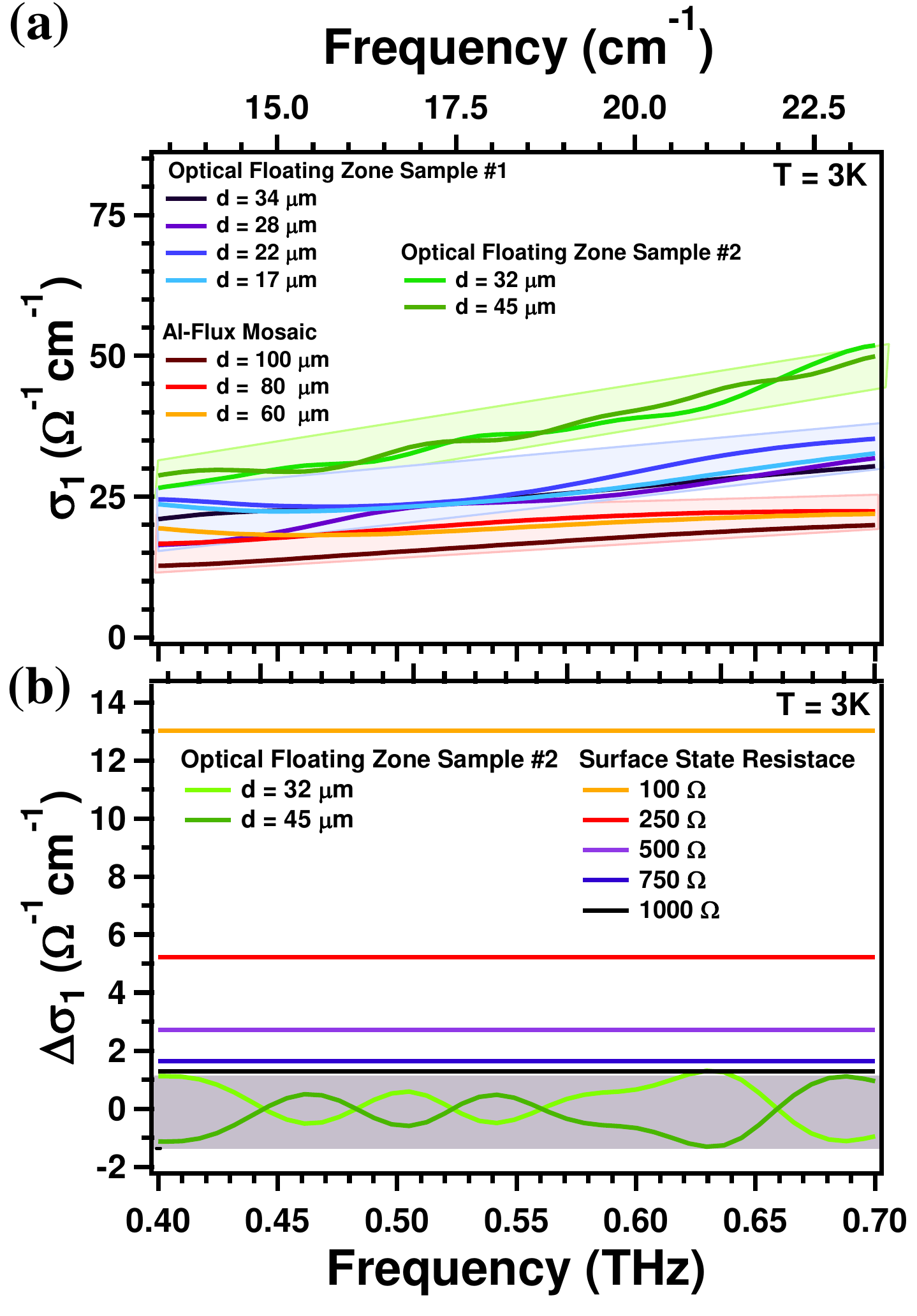}
\caption{(a) Thickness dependence of the optical conductivity at T = 3K $\pm$ 0.1K in the frequency range of the highest signal to noise of our measurement.  Data from two different optical floating zone samples are presented as well as data from the aluminum flux grown mosaic.  The colored regions represent the estimated experimental uncertainty of our measurement for each sample.  One can see that no systematic dependence on thickness is observed, indicating 3D bulk conduction. (b) Change in optical conductivity expected if surface states were present with the given sheet resistance as derived from our RefFIT tri-layer model.  The two lower curves are the conductivities from the optical floating zone sample \#2 presented in (a) with the average of the two conductivities subtracted.  The gray box represents our estimated measurement uncertainty.  From this we conclude that if surface states are present then they must have a sheet resistance  R$/\square \ge$ 1000 $\Omega$.  See text for details.}
\label{Fig3}
\end{figure}

As stated in the methods section above, the real and imaginary parts of the complex optical conductivity can be extracted from the complex transmission via numerical inversion of Eq. \ref{Transeq2}.  Figs. \ref{Fig2} (c,d) display the real part of the optical conductivity, $\sigma _1(\omega , T)$, extracted from the two transmissions shown in Fig. \ref{Fig2} (a,b) respectively.  With some notable differences to be discussed below, the general frequency and temperature dependence of these data are in rough agreement with those of previously reported optical studies \cite{Travaglini1984, Kimura1994, Nanba1993, Gorshunov1999}, although the exceptionally high resolution of our measurements provide new details.

A crossover from metallic to insulating behavior can be seen as a function of temperature in the conductivity of both samples, which show qualitatively similar behavior.  At the highest measured temperature, T = 20K, a Drude-like response can be seen as the optical conductivity is largest at the lowest frequencies and is a decreasing function of frequency thereafter.  The Drude-like response indicates the presence of free charge carriers in the conduction band.  As the temperature is reduced the magnitude of the Drude response correspondingly decreases, disappearing at T $\approx$ 12K, at which point the conductivity is nearly frequency independent out to 2 THz.  At lower temperatures, T $<$ 12K, the conductivity becomes an increasing function of frequency, displaying approximately linear behavior below $\approx$ 1 THz. This change in functional dependence of the conductivity with frequency signifies a shift to a new conduction mechanism. Above 1 THz the conductivity saturates and displays little dependence with temperature.  The frequency dependence of the conductivity will be further addressed in the discussion below.

\subsection{Thickness Dependence}

To further investigate these in-gap states, spectra were taken as a function of sample thickness.  A thickness dependent study was performed on three samples, two optical floating zone crystals and the aluminum flux mosaic comprised of 10 individual single crystals.  To obtain the thickness dependence, spectroscopy was performed, then samples were mechanically polished to a reduced thickness as measured by a micrometer, then spectroscopy was performed again, etc.  As the conductivity carries the dimension of (resistance $\times$ thickness)$^{-1}$, one would expect that the optical conductivity would display thickness dependence if significant surface conduction exists.  For samples with bulk 3D conductivity one expects there to be no thickness dependence of the conductivity as reducing the thickness also increases the resistance of the sample rendering the conductivity unchanged. Thus, transmission experiments performed in this fashion can separate surface and bulk conduction as has been done in Bi$_2$Se$_3$ \cite{Aguilar2012, Wu2013}.

Fig. \ref{Fig3} (a) displays the results of our thickness dependent study of the optical conductivity at T = 3K $\pm$ 0.1K, in the frequency range in which the highest signal to noise is achieved. However, we mention that our conclusions are not particularly dependent on this temperature or frequency range.  Thickness dependence of three samples are shown.  The colored regions are representative of the experimental uncertainty of our measurements which will be used for further analysis below.  One can immediately observe that there is no systematic dependence with sample thickness of the extracted optical conductivity within the uncertainty of our experiment.  We therefore conclude that the principal signal of the residual conductivity of the in-gap states stems from 3D bulk conduction.  

With the 3D nature of the optical conductivity within the gap established, we now discuss how this relates to the TKI prediction of SmB$_6$.  Our measurements are not able to exclude topological surface states residing within the bulk Kondo gap.  However, if surface states exist in the gap then they must have a conductance below the detection threshold of our measurement. Thus, an estimate of our uncertainty can be used to place limits on the potential surface state conductance.  To do so, the optical transmission was modeled in RefFIT \cite{Kuzmenko2005}.  The T = 3K conductivity of the optical floating zone sample \#2, shown in Fig. \ref{Fig3} (a), was chosen for the model as the thickness dependence on this sample possesses the lowest experimental uncertainty.  A tri-layer model of surface state - bulk - surface state was developed to model the transmssion. The bulk conductivity was given by that of the d = 32 $\mu$m sample shown in Fig. \ref{Fig3} (a).  The two surface states were modeled as two identical 2D Drude responses, in agreement with the surface states observed in Bi$_2$Se$_3$ \cite{Aguilar2012, Wu2013}.  We assume that the conductance of these states is constant as a function of frequency in our spectral range.  This is consistent with the $\approx$ 10 THz scattering rate determined in quantum oscillations experiments \cite{Li2014}. Therefore, the surface state conductance would manifest in the tri-layer model as a frequency independent offset to the conductivity when the thickness of the sample is varied.  

The results of the model are shown in Fig. \ref{Fig3} (b).  Shown at the bottom are the optical conductivity at T = 3K of the optical floating zone SmB$_6$ sample \#2 for thicknesses of d = 32 $\mu$m and 45 $\mu$m with the average of the two thicknesses subtracted.  The gray box indicates our approximate uncertainty in the experiment.  The horizontal lines demonstrate the expected offset in the effective optical conductivity that would be extracted if surface states with the specified resistances existed in addition to the bulk conductivity.  From the model we conservatively conclude that we would be able to identify surface states with a sheet resistance  R$/\square \le$ 1000 $\Omega$ in our experiment.   Therefore, surface states with a sheet resistance below this detection threshold can be excluded.

\subsection{Coupling of Bulk States To Spin Excitons}

The results presented above show that although SmB$_6$ may be a bulk dc insulator, it shows significant bulk ac conduction.  Low energy 3D bulk states exist within the gap of SmB$_6$.  These states within the gap can also greatly affect other low energy excitations of SmB$_6$.  For instance, a recent neutron scattering study proposed and observed a collective spin exciton at $\approx$ 16 meV which results from the electron-hole continuum \cite{Fuhrman2015, Fuhrman2014}.  The width of the exciton was observed to be exceptionally narrow, $\approx$ 2 meV, although more recent measurements with improved resolution suggest the width of the resonance to be even narrower than that, $\approx$ 100 $\mu$eV \cite{FurhmanPC}.  The narrow width suggests the spin exciton to be an extremely long lived excitation and was speculated to be protected from decaying into electron-hole pairs by the hybridization gap in which it resides. 

However, the exciton can in principle couple to states within the gap, whether they are topological Dirac states or bulk states.  Coupling of the spin exciton to such states can have tremendous impact on the physics of SmB$_6$.  The possibility of spin excitons coupling to surface states has been discussed theoretically \cite{Kapilevich2015} and then reported experimentally via tunneling spectroscopy \cite{Park2016}.  Meanwhile, evidence of coupling between spin excitons and bulk states was recently presented via Raman spectroscopy measurements \cite{Valentine2016}.  In that work it was found that disorder in the form of Sm$^{+2,3}$ vacancies on the order of only 1\% leads to states within the gap.  The spin exciton shows a corresponding spectral broadening with increasing disorder suggesting decay through these bulk in-gap states.

While the spin exciton lies at higher energy than what our experiment can access, we can still quantify how the finite density of states within the gap couples to these collective excitations.  A similar analysis has been performed previously in regards to how crystal field line widths in metallic systems and the ``resonance mode" in high T$_c$ cuprate superconductors \cite{Kee2002} couple to a continuum of states.  One may use the expression

\begin{equation}
\Gamma = 4 \pi [g V_c D(\epsilon _F)]^2 \Omega
\label{Coupeq}
\end{equation}

\noindent where $\Gamma$ is the full width at half maximum of the resonance, $\Omega$ is the exciton energy, D($\epsilon _F$) is the density of states at the Fermi level, and $g$ is the coupling constant.  

With the energy, $\Omega \approx$ 16 meV, and width, $\Gamma \approx$ 2 meV, of the spin exciton measured by neutron and Raman experiments \cite{Fuhrman2015, Valentine2016}, we can extract the coupling constant if the density of states at the Fermi level is known.  An estimate of the density of states can be obtained from the metallic contribution of the heat capacity, C$_\text{ele} = \gamma$T.  Interestingly, heat capacity measurements find a surprisingly large metallic component at low temperatures, often on the order of 10 mJ K$^{-2}$ mol$^{-1}$, in agreement with the observed large low energy spectral weight \cite{Flachbart2006, Gabani2001, Phelan2014}. Additionally, recent measurements indicate that the large metallic heat capacity is independent of sample surface area and is therefore of bulk origin \cite{Wakeham2016}.  Phelan \textit{et al.} report a value of $\gamma$ = 25 mJ K$^{-2}$ mol$^{-1}$ in optical floating zone samples \cite{Phelan2014}.  In the simplest picture of a non-interacting Fermi gas, the density of states is proportional to $\gamma$, in the units of the given heat capacity, as, $\gamma = \frac{\pi ^2  k_{B}^2 N_A V_c}{3} D(\epsilon _F)$, where $k_{B}$ is Boltzmann's constant, $N_A$ is Avagadro's number, and $V_c$ is the volume of one SmB$_6$ formula unit.  Substituting the observed value of $\gamma$ into this expression and then the corresponding density of states into Eq. \ref{Coupeq} results in a coupling constant of $g$ = 9.40 meV.

The coupling constant is more easily understood in the conventional dimensionless form, $\lambda$, which can be determined via the expression, $\lambda = \frac{2 I_0 g^2 V_c D(\epsilon _F)}{\Omega}$.  Here, $I_0$ is the ratio of the integrated spectral weight of the excition to the total integrated spin structure factor. An $I_0$ $\approx$ 0.4 was determined from neutron scattering experiments \cite{Fuhrman2015}.  Substituting in the appropriate values gives $\lambda$ = 0.047.  This calculation shows that the coupling of the excitons to the bulk in-gap states to be very weak.

The strong dependence of the exciton's linewidth on sample disorder \cite{Valentine2016,FurhmanPC} is interesting considering the relatively weak dependence of the in-gap states we probe.  Moreover, the fact that the exciton is seen clearly in Raman, whereas it it is not observed in the infrared \cite{Gorshunov1999} points to a well-defined selection rule  associated with its excitation.   In inversion symmetric systems like SmB$_6$ excitations are either Raman active or infrared active, but not both.   We therefore speculate that the exciton is - in the ideal case - prevented by symmetry from coupling to the infrared continuum and it is only through disorder that this coupling becomes finite.   In other words, the exciton's lifetime disorder dependence comes from a strong dependence of disorder on $g$ in Eq. \ref{Coupeq} and not $D(E_F)$.

\section{Discussion}

Our results and the existing heat capacity data show that the low energy density of states in SmB$_6$ is quite large, in contrast to the assumption of a clean insulating gap.  Why then do transport experiments claim to see a perfectly insulating gap with activated dc transport?  First, it is important to point out that in the limit of zero temperature dc transport can only probe extended states.  However, ac experiments are also sensitive to \textit{localized} quasiparticle states, as well neutral excitations that carry a dipole moment (e.g. phonons as the least exotic example).  In ac experiments, charge does not need to be transmitted across the sample as charge in localized states can still oscillate at ac frequencies on length scales smaller than the localization length and dissipate energy.  Samples which display such behavior can appear as insulators in dc transport experiments but conductors at finite frequency.  In this regard, we remind the reader that the ac conductivity in the THz range in SmB$_6$ is orders of magnitude greater than the dc value at low temperatures \cite{Travaglini1984, Kimura1994, Nanba1993, Ohta1991} in agreement with this picture.

What is the origin of the in-gap ac conduction?  The most obvious explanation is that it originates from impurity states.  A number of authors have pointed out the special role of impurities in Kondo insulators, which in some cases can form a Kondo hole impurity band \cite{Sollie1991, Schlottmann1992, Schlottmann1996, Riseborough2003}.  Yet, these scenarios predict magnetic phenomena which are not observed.  However, the general phenomenology of the low temperature ac and dc conductivity of SmB$_6$ is somewhat similar to what is observed in some localization-driven insulators, such as the disordered electron glass Si:P \cite{Helgren2004}.  In the latter systems the dc conductivity is described by a model of variable range hopping with a stretched exponential activated dependence and a power law dependence of the ac response.  The expected dependencies are determined by the form of the density of states at the Fermi level \cite{Mott1968,Helgren2004, Shklovskii1981}.  Assuming a nearly constant density of states, one expects the dc conductivity, for 3D hopping conduction to follow the expected Mott form for Fermi glasses going with temperature as $\ln(\sigma _\text{dc}) \propto T^{-\frac{1}{4}}$ \cite{Mott1968}.  Indeed Gorshunov \textit{et al.} claim such a temperature dependence for temperatures 4K $\leq$ T $\leq$ 10K with a characteristic energy scale of T$_0$ = 54K, although fitting an exponential to such a small range cannot be considered very conclusive.  In such insulators where disordered induced localization is expected to be central to the physics, the expectation is that at the lowest temperatures ac conduction occurs between $resonant$ pairs of localized states.   Without interactions the ac conductivity is expected to follow Mott's famous $\omega^2$ law, which is clearly inconsistent with the data exhibited here.   With interactions included, but at frequency scales below that of the characteristic interaction energy between electron-hole pairs, the expectation is that the conductivity is quasi-linear with $\sigma_1 (\omega)$ = $e^4  D(\epsilon _F)^2 \xi^4 [\mathrm{ln}(2I_0/ \hbar \omega)   ]^3 \omega / \epsilon$ where $\xi$ is the localization length \cite{Shklovskii1981}. $I_0$ is the characteristic scale of tunneling between localized states that is expected to be bounded by the hybridization gap energy.   Our ac conductivity data (Fig. \ref{Fig2}) is roughly consistent with this linear dependence at our lowest measured frequencies.  It is also important to point out that in principle, even the ``metallic" heat capacity seen in SmB$_6$ is consistent with localized states as it has been emphasized that despite their insulating nature such systems can still show a $fermionic$ linear in $T$ heat capacity  (albeit of a magnitude far less than observed in the present case as discussed below) \cite{Kobayashi77a,Wagner97a,Mael86a,Rogatchev00a}.

However, despite the (partial) qualitative agreement with a picture of localized bulk states, there are important  quantitative issues that need to be resolved.  For instance, the magnitude of the ac conductivity in the present case is quite unlike other disordered insulators.   It is approximately four orders of magnitude larger than both the impurity band conduction in Si:P (at say doped 39\% of the way towards the 3D metal-insulator transition) \cite{Helgren2004} and is essentially of the scale of the ac conduction in completely amorphous Nb$_x$Si$_{1-x}$ \cite{Helgren01a} alloys.  Although in principle this very large scale of the ac conductivity can follow from the very large $D(E_F)$ in SmB$_6$, the large heat capacity itself is unexplained.  Although localized states at $E_F$ can manifest a linear in T heat capacity, the heat capacity of SmB$_6$ is many orders of magnitude larger than any known localization-driven insulator  ($\sim 10 \mu$J K$^{-2}$ mol$^{-1}$ for the impurity band in Si:P at $\sim$ 50 \% of the $x_c$ for the metal-insulator transition \cite{Kobayashi77a,Wagner97a} and $\sim 0.5$mJ K$^{-2}$ mol$^{-1}$ for amorphous glasses \cite{Mael86a,Rogatchev00a}).   Additionally, localized states at $E_F$ will more generically result in stretched exponential variable range hopping style-activation and not simple activation in the transport.

Gorshunov \textit{et al.} claim that sample imperfections manifests as a slight maximum in the real conductivity at 0.72 THz (24 cm$^{-1}$) \cite{Gorshunov1999}.   Although, this is an energy scale that matches the activation energy scale of the dc resistivity above 10K, we have observed no such band in any sample measured in this study.   Moreover, Gorshunov \textit{et al.}'s band was only a weak maxima, and it is not clear (even if such a band was present) why it would manifest in the dc data with an activated temperature dependence.  It has also been found that the activation energy is strongly dependent on pressure \cite{Cooley95a}, which has no obvious explanation where the activated transport arises through hopping in an impurity band.

Therefore, one should consider the possibility that these in-gap states are intrinsic to SmB$_6$.  The apparent agreement in the optical conductivity in our measurements between samples grown by different methods and under varying conditions suggests a different explanation than impurities.  One can see from Figs. \ref{Fig2} and Fig. \ref{Fig3}(a) that the low temperature conductivities of the samples measured in this study vary by - at most - a factor of 2.  Generally, insulating states induced by disorder have conductivities that are exponentially sensitive to the degree of impurities, often displaying large variation in physical properties upon even small changes to the sample composition.  The apparent lack of dependence upon sample preparation and, in some cases, doping \cite{Phelan2014} suggests the intrinsic nature of these localized states.  We remind the reader that the aluminum flux grown mosaic was comprised of 10 individual single crystals and is therefore likely representative of samples grown by this method. 

A number of possibilities exist for ac conduction by an intrinsic mechanism at low energy.  One theory suggests that a Fermi surface comprised of electrically neutral quasiparticles can exist within the Kondo gap \cite{Coleman1993, Baskaran2015}.  These quasiparticles, although electrical neutral, may still possess an electrical dipole moment and therefore conduct at ac frequencies \cite{Ng07a}.  A separate theory claims that these in-gap localized states may originate from intrinsic electrons in SmB$_6$ that become self trapped through interactions with valence fluctuations \cite{Curnoe2000}.  Additionally, a recent torque magnetometry experiment has claimed to observe unconventional quantum oscillations stemming from a bulk 3D Fermi surface in SmB$_6$ \cite{Tan2015}.  These results suggest that the potentially intrinsic nature of our observed in-gap localized states warrants further consideration and investigation.

Lastly, we discuss the limits placed on the potential surface state sheet resistance from our data.  As discussed above,  Fig. \ref{Fig3} (b) demonstrates that the surface states of the SmB$_6$ samples studied must have a sheet resistance of R $\ge$ 1000 $\Omega$ or they would be detectable in our measurement.  The reported sheet resistance of surface states in SmB$_6$ varies greatly between transport experiments, ranging from 0.1 - 100 $\Omega$ \cite{Kim2013, Wolgast2013, Chen2015}. This wide discrepancy may originate from the unknown conduction paths in 4 probe measurements as current can in principle travel along all surfaces of the sample or perhaps from differences in surface preparation methods.  A benefit of our optical experiments is that the conduction paths are precisely known as the measurement geometry is well-defined.  Correspondingly, larger values of surface state sheet resistance are often reported from optical techniques such as R = 250 $\Omega$ \cite{JZhang2015} in SmB$_6$ thin films and R $\approx$ 200 $\Omega$ in Bi$_2$Se$_3$ \cite{Aguilar2012, Wu2013}.  It is unclear if the mechanical polishing performed on the SmB$_6$ samples in this study can account for such a discrepancy in reported sheet resistance.  However, we point out that while the floating zone single crystals were mechanically polished on both front and back surfaces, the aluminum flux grown samples present their as-grown surface on one side.  If high mobility surface states existed then they would be presumably maintained on this surface of these samples and observed in our experiment.   Moreover, we point out that a recent study which investigated the effects of polishing on surface state resistance found that fine polishing \textit{increased} surface resistance as it removed conductive subsurface cracks in the sample \cite{Wolgast2015}.  Correspondingly, the observed surface resistance on highly polished samples was found to be 2-3 k$\Omega$ through surface sensitive Corbino measurements, in agreement with the R/$\square$ $\ge$ 1 k$\Omega$ limit found in this study \cite{Wolgast2015, Kurdak2016}.

\section{Conclusion}

In this work we presented a detailed study of the optical properties of SmB$_6$ in the THz frequency range.   SmB$_6$ single crystals grown by both optical floating zone and aluminum flux methods were studied and found to be consistent in their optical properties.  We show, through high resolution time domain terahertz measurements, that there is substantial in-gap 3D bulk ac conductivity in SmB$_6$.  We discussed the possible origins of these states and their coupling to the low energy spin excitons of SmB$_6$ in which a coupling constant of $\lambda$ = 0.047 was found.  A modeling of the optical conductivity concluded that any potential surface states, which must lie below our detection limit if present, must have a sheet resistance of R$/ \square$ $\ge$ 1000 $\Omega$, which is substantially larger than what has been previously reported.   Our results demonstrate the hybridization gap of SmB$_6$ is insulating in dc transport measurements but in fact displays significant bulk conduction at finite frequencies.

\section{Acknowledgments}

Work at the Institute for Quantum Matter (IQM) was supported by the U.S. Department of Energy, Office of Basic Energy Sciences, Division of Materials Sciences and Engineering through Grant No. DE-FG02-08ER46544.  NJL acknowledges additional support through the ARCS Foundation Dillon Fellowship. TMM acknowledges support from The David and Lucile Packard Foundation. ZF acknowledges support from NSF Grant No. DMR 08-01253. Work at UCI was also supported by Grant No. FAPESP 2013/20181-0. Work at Los Alamos National Laboratory (LANL) was performed under the auspices of the U.S. Department of Energy, Office of Basic Energy Sciences, Division of Materials Science and Engineering. P.F.S.R. acknowledges a Director's Postdoctoral Fellowship through the LANL LDRD program.  We would like to thank P.-Y.  Chang, P. Coleman, N. Drichko, O. Erten, W. Fuhrman, C. Kurdak,  P. Riseborough, F. Ronning, S. Sebastian, and M. E. Valentine for helpful conversations.

\bibliography{SmB6Bib}

\end{document}